\documentclass{elsart}
\usepackage{graphicx}
\usepackage{dcolumn}
\usepackage{bm}
\usepackage{amsmath}
\usepackage{comment}

\usepackage{color} 

\newcommand{\chem}[1]
{
\ensuremath{\mathrm{#1}}
}

\begin{document}
\begin{frontmatter}

\title{Demonstration of a 2x2 programmable phase plate for electrons}

\author[emat]{Jo Verbeeck}
\ead{jo.verbeeck@uantwerp.be}
\author[emat]{Armand B\'ech\'e}
\author[emat]{Knut M\"uller-Caspary}
\author[emat]{Giulio Guzzinati}
\author[cea]{Minh Anh Luong}
\author[cnrs]{Martien Den Hertog}
\address[emat]{EMAT, University of Antwerp, Groenenborgerlaan 171, 2020 Antwerp, Belgium.}
\address[cea]{Univ. Grenoble Alpes, CEA, INAC, MEM, LEMMA Group, F-38000 Grenoble, France}
\address[cnrs]{Institut Neel, 25 avenue des Martyrs BP 166, 38042 Grenoble Cedex 9, France.}

\begin{abstract}
First results on the experimental realisation of a 2x2 programmable phase plate for electrons are presented. The design consists of an array of electrostatic einzel lenses that influence the phase of electron waves passing through 4 separately controllable aperture holes. This functionality is demonstrated in a conventional transmission electron microscope operating at 300~kV and results are in very close agreement with theoretical predictions. The dynamic creation of a set of electron probes with different phase symmetry is demonstrated, thereby bringing adaptive optics in TEM one step closer to reality. The limitations of the current design and how to overcome these in the future are discussed. Simulations show how further evolved versions of the current proof of concept might open new and exciting application prospects for beam shaping and aberration correction.
\end{abstract}


\end{frontmatter}

\section{Introduction}
Adaptive optics, the technology to dynamically change the phase transfer of optical elements has its roots in astrophysics, where dynamically changing telescope mirrors can compensate for time varying atmospheric induced aberrations for optimised observation from earth \cite{fried_optical_1966,davies_adaptive_2012,wang_optical_1977,merkle_successful_1989}, as well as in space \cite{clery_building_2016}. The technology has greatly improved since and has sparked an avalanche of innovative uses in many different areas where dynamic control over optical elements is wanted. Examples are endoscopic imaging through a mulltimode optical fiber \cite{mahalati_resolution_2013}, second harmonic imaging as e.g. used in stimulated emission depletion microscopy (STED) \cite{hell_breaking_1994}, focusing inside thick semitransparent objects and tissue \cite{booth_adaptive_2014}, optical quantum encryption \cite{clemente_2010}, laser welding \cite{mrna_2013} and many more. All these breakthroughs were made possible by the existence or development of so-called spatial light modulators, devices that allow for a programmable change of the phase of optical waves when passing through or being reflected from the modulator. Different designs exist based on liquid crystals \cite{slinger_computer-generated_2005,igasaki1999}, piezo controlled mirror segments \cite{steinhaus_1979}, electrostatic or magnetic influencing of coated flexible membranes \cite{vdovin_1995}, shape changing flexible refractive elements as in mamal eyes \cite{hong_2006} and many more. Each technology has its own advantages and disadvantages in terms of insertion losses, speed, constraints to what phase functions are allowed, pixel count, power handling of intense light beams, precision and so-on.
In TEM, on the other hand, adaptive optical elements are commonplace. Indeed, even the simplest round magnetic or electrostatic lens, is tunable by either changing the current through the coils or by changing the potential difference over electrodes \cite{rosebook,hawkesbook}. Adaptive optics is readily available on modern aberration corrected (S)TEM instruments in the form of complicated assemblies of multipolar lenses. They can be adaptively optimised on test samples to obtain the vastly improved resolution and current density that has formed the basis of most of the successes in experimental electron microscopy over the last decade \cite{haider_current_2009,tromp_new_2010,batson_sub-angstrom_2002,rose_prospects_2005,rose_phase-contrast_1974,linck_chromatic_2016,haider_electron_1998}.
Yet, aberration correctors in their current state don't allow for full flexibility in the phase they imprint on the electron wave. This is very apparent from the many attempts that have been, and are being made, towards a so-called Zernike phase plate for optimising the contrast in (weak) phase objects \cite{zernike_phase_1942}. Indeed, if an aberration corrector would be able to change the phase on the optical axis by a given amount while leaving the rest of the wave unaffected there would be no need for the veritable zoo of different phase plate designs that exist. The magnetic vector potential in a magnetic multipole corrector is determined by the individual poles that act as boundary conditions to the free space in which the electrons travel. The vector potential component in the direction of electron motion obeys a Poisson equation:
\begin{equation}
\nabla^2 A_z -\frac{1}{c^2} \frac{\partial^2 A_z}{\partial^2 t}=-\mu_0 J_z,
\end{equation}
assuming straight trajectories along z-direction through a thin region of space where the vector potential is non-zero. The electron, due to the Aharanov-Bohm effect, gets phase shifted by the projected magnetic vector potential $A_z$ along its path as $\phi(x,y)=\frac{q}{\hbar}\int_{-\infty}^{\infty} A_z dz$. Assuming time independent solutions in free space ($J_z=0$), we note that the obtainable phase plates are limited to a (small) subset of all possible phase plates constrained by a Laplace equation in 2D:
\begin{equation}
\nabla^2_{x,y} \phi = -\frac{q}{\hbar}\frac{\partial A_z}{\partial z}\Big|^{\infty}_{-\infty}=0.
\end{equation}
For thicker lenses (all realistic cases), the trajectory can deviate from the z-direction and more complicated arguments are needed, but in practice, for reasonable boundary conditions or multipole orders, the available phase plates tend to be smooth. Fortunately, the available phase maps advantageously to the Zernike polynomials which are closely linked to the typical low order aberrations that are dominant in the electron microscope. However, a sharp change in phase in the center of the field as required for e.g. a Zernike phase plate would require prohibitively large magnetic multipole orders and is impractical for the foreseeable future.

To overcome this limitation, the assumption of free space has to be dropped (or alternatively dynamic fields are needed) and all current Zernike phase plates indeed consist of some form of material that is in contact with the electron wave. Some of the more promising designs create a miniaturised electrostatic einzel lens or the Zach variant in order to obtain a tuneable local phase shift. This comes at the expense of admitting the material lens in the beam path resulting in a partial loss of electrons, decoherence, inelastic losses and charging issues \cite{hettler_high-resolution_2016,schultheiss_new_2010,frindt_-focus_2014}. A commercially available alternative, the Volta phase plate, uses local charge buildup on an insulating electron transparent film; but also here the free space requirement is dropped resulting in possible artefacts like drift, (in)elastic scattering, decoherence, uncontrolled charge and discharge \cite{danev_cryo-em_2016,danev_practical_2009,danev_single_2008,danev_volta_2014}.
So far, most phase plates have focused on phase contrast improvement and typically consist of a single region in space that is shifted in phase with respect to the rest of the wave that is left mostly unaltered. Increased flexibility in phase manipulation of electron waves recently gained considerable attention, showing exotic phase profiles as in electron vortices \cite{Bliokh2017,McMorran2011,Uchida2010,Verbeeck2010,beche_highres_vortex,edgecomb2017}, Airy waves \cite{voloch-bloch_generation_2013,Kaminer2015,Guzzinati2015}, helicon beams \cite{shiloh_3d_2017} or even the creation of an institute logo by modulating the phase profile of the wave \cite{shiloh_sculpturing_2014}. Nearly all of these experiments make use of  either holographic reconstruction using specifically crafted phase \cite{Grillo2014,grillo_generation_2014,rubinsztein-dunlop_roadmap_2017} and or amplitude gratings \cite{Verbeeck2010,McMorran2011,Verbeeck2011b} or refractive elements crafted with specific height profiles \cite{Uchida2010,Grillo2014,beche_fib_phaseplate,shiloh_spherical_2017,grillo_2017_cscorrect}. Even though these methods provide unprecedented flexibility in phase programming, offering a rich resource for the exploration of e.g. non-diffractive optical modes, they suffer all from being static techniques. As such, a considerable time is required to create such phase plates and swap them in the aperture position of existing TEM microscopes which needs to be repeated with any new phase plate that is needed.

In this paper we expand on these ideas by creating a rudimentary proof of concept of a programmable phase plate consisting of an array of 2x2 einzel lenses, offering full dynamic control over the 4 individual phase plate elements. Even though the design has clear drawbacks in terms of total transmissivity, and limited pixel count, it nevertheless constitutes an important step towards a more generic programmable phase plate for TEM. In the remainder, we will give details on the design and fabrication of this phase plate followed by the first experimental results proving that a programmable multi-element phase plate for electrons is feasible. We continue by discussing the uses of such a limited phase plate and speculate about how much upscaling would be required to enable novel and exciting methods in the TEM \cite{Guzzinati2015}.
It has to be noted here, that several alternative physical principles exist that can be exploited to create similar or even better devices, but we focus here on the description of what we believe to be the first truly versatile programmable phase array for electrons.

\section{Experiment}
\begin{figure}
    \includegraphics[width=\columnwidth]{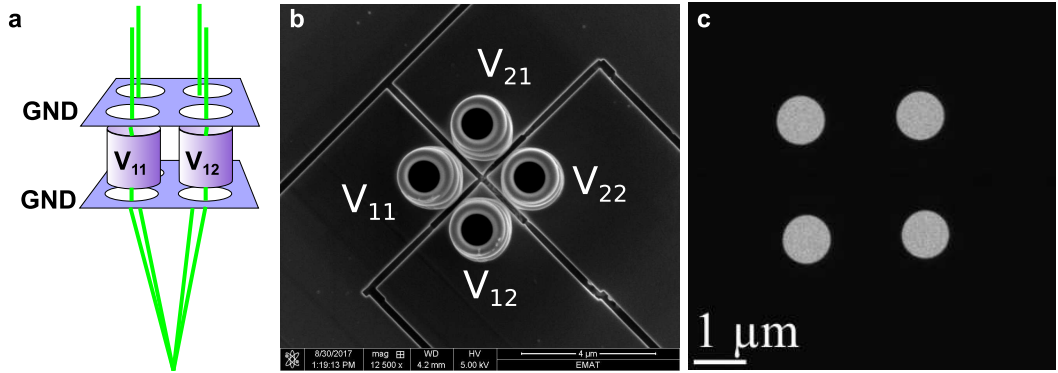}
    \caption{(a) sketch of an array of weakly excited electrostatic einzel lenses individually controlling the phase of 4 electron beams which recombine in the far field to form a programmable interference pattern. (b) SEM image of a proof of concept implementation of a 2x2 phase plate prepared as described in this paper. (c) TEM image of the aperture showing the geometry of the 4 aperture holes as well as the limited fill factor. }
    \label{setup}
\end{figure}

In order to prepare an array of electrostatic einzel lenses as sketched in fig.~\ref{setup}.a,  a
set of cylindrical electrodes needs to be created, sandwiched between two ground planes. Here we
simplified the setup considerably by working with a single large ground plane on which four
independent cylinder electrodes are placed. The omission of the top ground plane will lead to a
minor leaking of the potential of one cylinder into the space above a neighbouring cylinder
electrode. This effect results in a small amount of crosstalk between the electrodes which can
be partially compensated by altering the individual potentials to obtain the desired phase
profile. Note that the name einzel lens can be confusing here, as the lensing function of these
lenses is deliberately kept so low that the dominant effect is a pure phase shift depending on the
total projected potential. This also means that at the low voltages that are being used, possible aberrations caused by the lenses are negligible compared to the phase shifting effect .

\begin{figure}
    \includegraphics[width=\columnwidth]{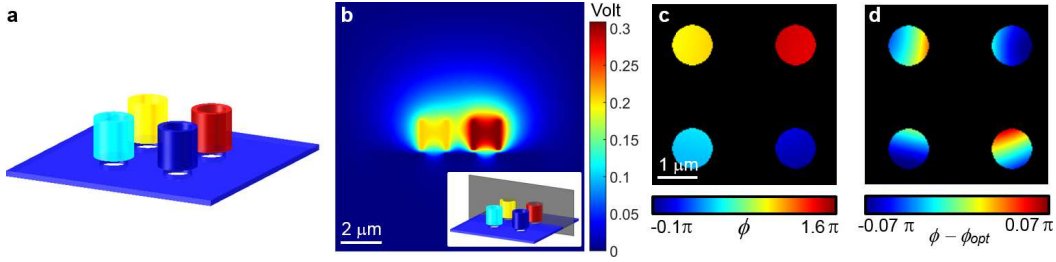}
    \caption{Simulated potential of the setup (a) showing the four electrodes at different potential of $0, 103, 206, 309$~mV in order to obtain $0,1/2\pi,\pi,3/2\pi$ phase shift. A slice through 2 neighbouring electrodes (b) shows the potential homogeneity inside the cylinders with fringe fields extending above the electrodes due to the missing top ground plane. The projected potential is shown in (c) in units of phase shift assuming 300~kV electrons. The difference with the intended phase plate of $0,1/2\pi,\pi,3/2\pi$ is shown in (d) with a maximum deviation that remains below $0.07\pi$ as a result from the cross talk between the neighbouring pixels due to the lack of a top ground plane. }
    \label{potential}
\end{figure}

A numerical solution of the Poisson equation for the half-einzel lens setup is presented in fig.\ref{potential} making use of a Liebmann algortihm and shows the four electrodes at different potential in fig.\ref{potential}a. A cross section through 2 neighbouring electrodes in fig\ref{potential}b shows the fringing fields above the electrodes which lead to some cross talk. The projected potential is shown in fig.\ref{potential}c expressed in units of phase assuming 300~kV electrons and with a target phase of $0,1/2\pi,\pi,3/2\pi$ for each electrode respectively. The relative deviation from the intended flat phase in each electrode is plotted in fig.\ref{potential}d showing the cross talk effect leading to phase errors of less than $0.07\pi$ in this case. Future geometrical optimisation could reduce this effect further or alternatively a top ground electrode would make this effect negligible at the expense of a more complicated setup.

The device fabrication starts from a commercial \chem{Si_3N_4} TEM sample grid with a 200~nm 250$\times$250
$\mu$m$^2$ square membrane. The grid was coated on both sides with about 500~nm of gold, making
use of a Balzers SCD 004 sputter coater. Masks were used to make sure the top and bottom layer are not electrically contacted. In the next step, four platinum cylinders of 1.4~$\mu$m  height were
deposited on the top gold surface by focused ion beam induced deposition (FIBID) using an FEI Helios
Nanolab. The cylinders are placed on a square grid with a distance of approximately 2.5 $\mu$m from
center to center. The four cylinders were then electrically insulated from each other by FIB
milling the top layer of gold until reaching the underlying SiN membrane. Thanks to the large
atomic number difference between the gold and the SiN membrane,it was possible to mill proper
electrodes without milling through the SiN membrane, ensuring good insulation between the bottom
ground electrode and the four top electrodes. As a final step, 1 $\mu$m circular holes were
drilled in the Pt cylinders all the way through the stacked layers to let the electron beam pass.
The 1 $\mu$m diameter was chosen in order to have an aspect ratio between diameter and height of the cylindrical electrodes of approximately 1. This provides a rather homogeneous potential profile inside the tubular electrode as can be judged from the electrostatic simulation in fig.\ref{potential}.

The device was placed in a Dens Solutions Wildfire in-situ heating chip holder on top of a
sacrificial heating chip, only used for contacting (heater window mechanically removed). The final
contacts from the heating chip to the SiN grid are made with silver paint. Then, conductivity
between the electrodes or between the electrodes and ground were carefully checked with a Keithley
2400 Source Meter and possible short circuits are corrected by further ion milling if necessary. The device was finally placed
under vacuum conditions to allow for outgassing prior to insertion in the transmission electron
microscope.

Once the contact holder was inserted in an aberration corrected FEI Titan$^3$ microscope (operating at 300~kV), the
assembly was connected to a set of tuneable voltage sources. The microscope was operated in
Lorentz mode with a strongly defocused monochromator in order to obtain sufficient spatial
coherence in the sample plane to cover the four einzel lens elements. The apertures were then
illuminated with a very good approximation of an electron plane wave. By going in diffraction
mode, it was possible to observe the far field pattern of the four apertures. The diffraction
pattern was recorded both in focus and defocused by 500~nm to better visualize the interference formed by the four individual electron wave patches that travel through the four aperture holes.

\begin{figure}
    \includegraphics[width=\columnwidth]{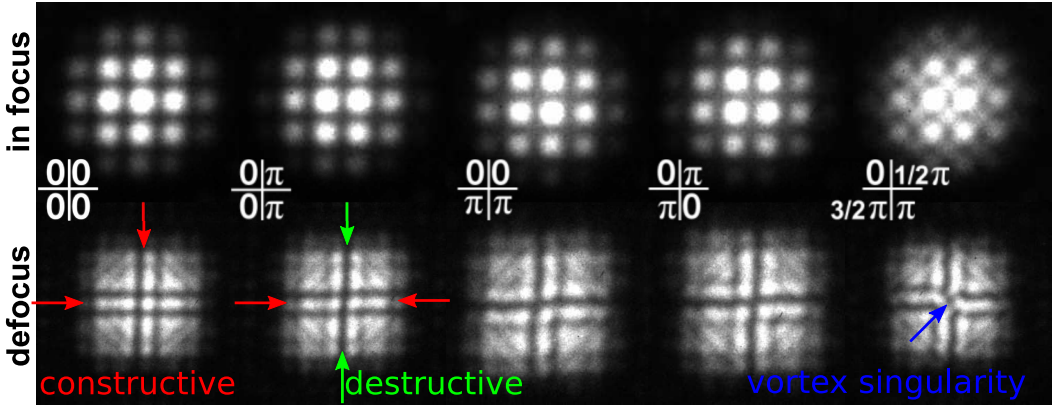}
    \caption{ (top row) Set of diffraction pattern for 5 different applied potential configurations leading to an approximate realisation of the phase symmetry as indicated on each panel. (bottom row) Defocused diffraction patterns for the same potential configuration showing more directly the constructive and destructive interference features between the 4 patches  as indicated by arrows.
    }
    \label{experiment}
\end{figure}

When all electrodes are grounded, the pattern displayed in fig.\ref{experiment} is obtained. One
can observe a clear constructive interference between the four patches in the defocused case.
Another pattern was obtained by applying a voltage to the two right electrodes until destructive
interference between the left/right patch occurs. This requires approximately 300~mV with some
deviation over the different electrodes due to remaining leakage current issues between the
electrodes and the height of each cylinder not being exactly the same. This demonstrates that indeed the pixel elements influence the phase of the electron wave by $\pi$. It is
then possible to estimate the sensitivity of the phase to the potential with a back-of
the-envelope calculation making use of the interaction constant of $\sigma=6.5~V^{-1} \mu m^{-1}$
for 300~kV electrons and a tube length of 1.4~$\mu$m. For a $\pi$ phase shift, this would require 350~mV which is in approximate agreement with the experimental values. A more accurate estimate can be made with the potential simulation presented in fig.\ref{potential} and leads to an estimated 205~mV needed to obtain pi phase shift, more accurately taking into acount potential inhomogeneities and fringe fields. The result however depends in an sensitive manner on the exact geometry which only approximately matches the actual shape of the device.

\begin{figure}
    \includegraphics[width=\columnwidth]{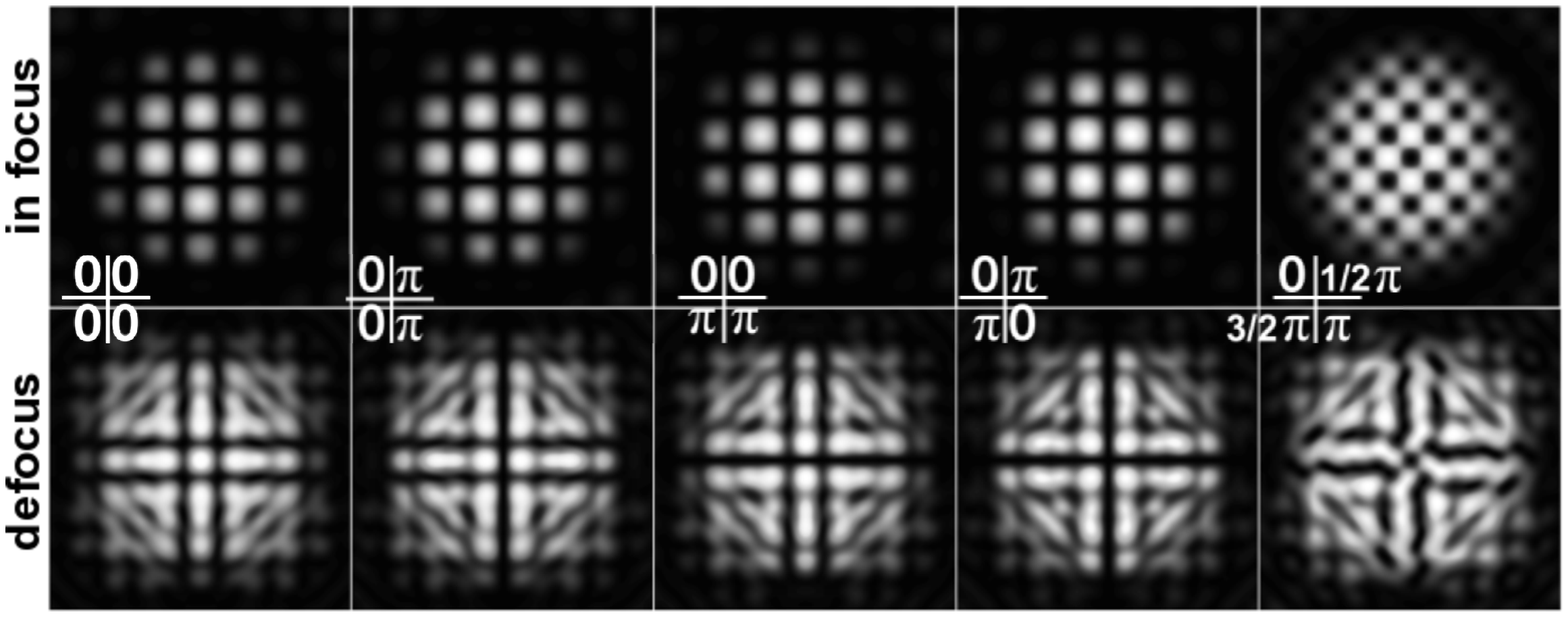}
    \caption{Simulated diffraction patterns assuming an ideal programmable phase shifting device with dimensions matching the experiment. Note the strong similarity in the intensity patterns for both in focus (top row) and defocused diffraction pattern series.
    }
    \label{simulation}
\end{figure}

 In a similar manner, a voltage difference
between up and down electrodes can be applied. As expected, the interference pattern has now up
down symmetry with a destructive interference between the up down patches. A quadrupolar pattern
and vortex pattern can also be generated, proving that a functional 2x2 programmable phase plate
has been created. The obtained interference patterns very closely match the simulated patterns given in fig.\ref{simulation}, assuming a true flat programmable phase plate device.

\section{Discussion}
\begin{figure}
\includegraphics[width=\columnwidth]{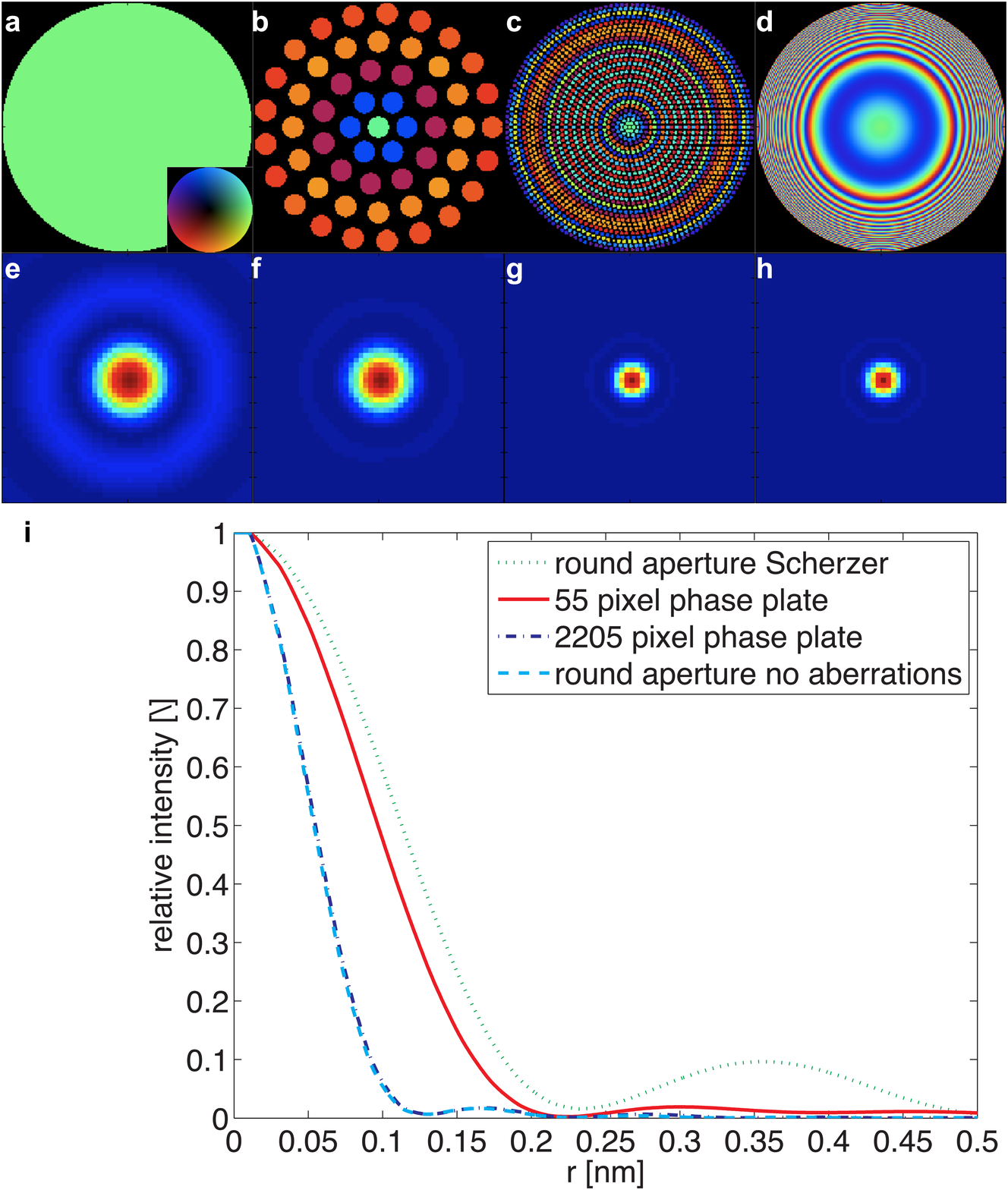}
\caption{Simulated performance of a programmable phase plate as a $C_s$ corrector assuming 300~keV and $C_s$=1~mm. Two phase plates are simulated, a modest 55 pixel plate and a more demanding 2205 pixel phase plate. The defocus of the lens system was optimised to -75~nm and -250~nm respectively to reach the optimum probe shape. The result is compared
to an ideal aberration free system with the same round aperture with opening angle 20~mrad. The resulting probe intensities
are shown respectively in (e,f,g,h), the images have dimension 1x1~nm. The azimuthally averaged probe intensity is shown in (i). Note that the low pixel count phase plate in (b) does not significantly improve the probe size but does improve the tails. Introducing significantly more pixels (c) approaches the ideal case quite closely. Note that even though the probe appears similar to the ideal diffraction limited case (d), a significant amount of the probe intensity is scattered to higher angles due to the narrow pixel shape function which would lead to an increased background when using this probe for imaging. The colorwheel shown as inset in (a) shows the colorscale used in (a,b,c,d) to depict both amplitude as intensity and phase as hue.}
\label{corrector}
\end{figure}

\begin{figure}
\includegraphics[width=\columnwidth]{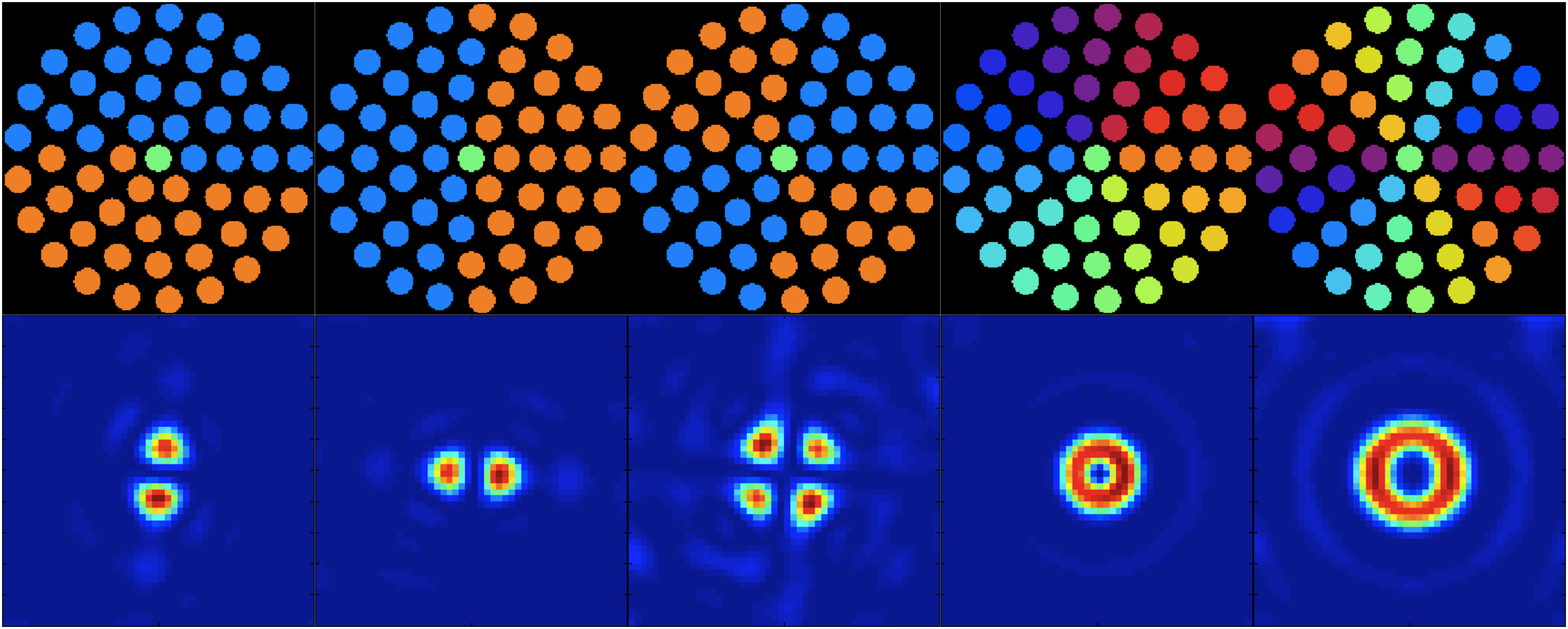}
\caption{A series of exotic electron beams prepared with a modest programmable phase plate (55 pixels, 41\% fill-factor) approximating (from left to right) $HG_{0,1}$, $HG_{1,0}$, $HG_{1,1}$, $LG_{0,1}$, $LG_{0,2}$. This shows that even a modest programmable phase plate can be very useful for producing exotic beam types that where hithertho difficult to produce in a TEM requiring time consuming and static (holographic) phase plates.}
\label{exotic}
\end{figure}

The above experimental results demonstrate the feasibility of a multi-element programmable phase plate for use in electron microscopy.  This holds clear promise for future work but several shortcomings of the current implementation will need to be addressed. The most important shortcoming lies in the inherent material making up the pixel element electrodes, blocking part of the electron beam. In the current demonstration this allows only about 12\% of the incoming beam to pass through the 4 individual lenses. Undoubtedly this so-called fill-factor can be increased substantially in later designs, but it is hard to imagine a design which would allow passing substantially more than 50\%. The fill-factor will of-course depend heavily on the micro machining or lithographic capabilities that will be used in further iterations of the design. As long as the phase plate is used in setups that shape the electron beam before the sample, this does not have to be a significant drawback, as modern instruments often provide more current or electron dose than the sample can handle, and losing a fraction of this current would not limit the usability of the device. Changing the phase behind the sample, as typical in e.g. setups for Zernike weak phase imaging, would suffer significantly from a less than unity transmitivity and it seems rather unfavourable to block electrons that carry precious information on beam sensitive materials. If however, this loss of electrons is compensated by a greater gain in contrast, it could still turn out beneficial. The blocking nature of the einzel lens structure shares however the significant drawback with other electrostatic Zernike phase plate designs that it throws away important low frequency information when used in the back focal plane, after the sample \cite{schultheis_fabrication_2006,schultheiss_new_2010}.
In order to upscale the device to a higher pixel count, lithographic techniques will be required and interconnect density may quickly put a limit to the maximum attainable number of pixels that each need to be individually contacted to a programmable voltage source. Matrix adressing methods such as those used in dynamic random acces memories could solve this, but require a nonlinear switching element, such as e.g. a transistor, integrated in the vicinity of each pixel element. Implementing such schemes should allow for pixel arrays of thousands to millions of pixel elements. One could wonder how many elements would be needed to provide ultimate flexibility in phase shaping the electron beam. As a demonstration, we compare the usefulness of a modest 55 and more elaborate 2205 pixel phase plate to act as a probe-$C_s$ corrector device in fig~\ref{corrector}. We assumed a spherical aberration coefficient of $C_s=1$~mm at 300~keV and compare to an uncorrected system at Scherzer defocus and to a fully corrected system, fixing the opening (half) angle at 20~mrad.
From the simulation we see that the 55 pixel variant shows some benefit in terms of reduced tails on the probe intensity (b,f) which results in a sharper intensity radial profile in fig.\ref{corrector}.i as compared to the uncorrected situation. Increasing the pixel count to 2205 results in a much sharper probe with a very simmilar profile (c,g)  as the diffraction limited profile of the fully corrected situation (d,h). This demonstrates that such phase plates could eventually start replacing magnetic multipole aberration correctors in the probe-forming system of TEM microscopes, but clearly a substantial amount of pixels would be needed. A full exploration of the ideal position, shape and number of pixels for aberration correction with a programmable phase plate will provide a rich arrea of further research, but is beyond the scope of the current discussion.
Nevertheless, aberration correction was not the first goal of the current implementation and low pixel element devices like the one presented here, or slightly upscaled can serve as a very useful experimenting platform for beam shaping and exploring other applications that focus more on tuning the phase of the electron wave to bring out specific contrast. Examples include the study of non-diffracting electron beams\cite{berry_nonspreading_1979,grillo_generation_2014-1,lapointe_review_1992,morris_realization_2010,mcgloin_bessel_2005,siviloglou_observation_2007,voloch-bloch_generation_2013}, symmetry mapping of plasmonic excitations\cite{guzzinati_probing_2017}, mapping of magnetic fields \cite{Verbeeck2010,Clark2014,Juchtmans2016,grillo_magnetic,schachinger_2017,Schattschneider2014} or edge contrast enhancement \cite{Juchtmans2016}. In optics, very useful spatial light modulators exist consisting of only 37 elements \cite{manzanera_2011}, even though they often provide a linear phase ramp in each pixel element which can also be imagined for electrons with a slight complication of the design where each einzel lens electrode is split in 3 separate planar electrodes providing a linear potential profile in triangular or hexagonal pixel elements. The optimised pattern of pixel elements depends on the intended use, but in low pixel count designs, it is likely that some form of radial pattern will be the most efficient.
In order to demonstrate some of the capabilities that e.g. a radially oriented 55 pixel phase plate would bring, we show in fig.~\ref{exotic} the capability of such a simple phase plate to create approximations to some well-known Hermite Gaussian and Laguerre Gaussian modes that are abundant in optics research. This capability would open up the field of beam shaping TEM providing a very desirable flexibility in the quantum state of the electron probe, much like what current spatial light modulators offer in optics. The fact that such a phase plate could rapidly switch between such modes could allow for differential measurements bringing out the specific contrast that comes with a certain probe symmetry and directly compare it to a near-simultaneous experiment done with another probe symmetry. As such, this development could bring the field of beam shaping to a significantly higher level.

Given the above demonstration that an upscaled version of the presented 2x2 device could have multiple uses, we can look into possible obstacles that could hinder progress. Indeed, the presence of the pixel electrodes can have unwanted effects, such as charging or decoherence due to thermal current flowing in the electrode material \cite{uhlemann_thermal_2015,uhlemann_thermal_2013,howie_addressing_2014}. The current implementation did not show either of these effects to be observable, but increasing the pixel density could eventually make these effects more prominent. In this respect, the short length (1.4~$\mu$m) over which the electrons interact with the pixel electrodes, helps to limit the decoherence effect substantially as they are expected to scale with interaction length and inversely with the square of tube radius \cite{uhlemann_thermal_2013}. Cooling could further help to keep decoherence low, but is rather unattractive from a practical point of view.
Contamination, could be another limitation of the design as small amounts of contamination could block pixels and render the whole array useless. This could be prevented or overcome by either including a heating element and or by allowing for easy replacement of a mass produced phase plate chip whenever this situation occurs. 

An important advantage of the presented device is the relative insensitivity of the performance to the quality of the voltage sources driving the pixel electrodes. Indeed in the current dimensions, a voltage of the order of a few 100~mV suffices to create a $2\pi$ phase shift. As phase is defined modulo $2\pi$ and because neighbouring patches of electron waves are divided by opaque walls of the pixel elements, there is no need to apply for more phase shift than this, similar to an optical Fresnel lens. This leads to a very relaxed requirement on the quality and noise performance of the voltage source. Indeed, even an idealised 8 bit digital to analog converter could provide a more than sufficient $2\pi/256$ phase resolution. This relative insensitivity of the phase to the potential also allows for very fast settling times in combination with a very low capacitance of the pixel elements with respect to each other and to the ground plane. As long as all pixels are driven from individual voltage sources (DA converters), a response speed in the range of micro to nanoseconds seems entirely feasible. Matrix addressing might limit this speed considerably depending on the design. Such rapidly changing phase plates could allow for autotuning and iterative measurement schemes which profit from the total absence of hysteresis effects that hamper all ferromagnetic core based optical elements in conventional TEM instruments.

\section{Conclusion}
We have demonstrated a proof of concept device that allows to dynamically control the phase of 2x2 segments of a coherent electron beam in a transmission electron microscope. The experimental implementation and first results demonstrate that such a device holds promise for upscaling towards a more useful higher number of pixels. Several design considerations and directions for further research are discussed. These make plausible that the presented proof of concept marks just the beginning of an exciting development that could alter the way how one thinks about electron optics, providing vastly increased flexibility, speed, repeatability and offering novel iterative measurement protocols that are difficult if not impossible to implement with current electron optical technology.

\section{acknowledgments}
J.V. and A.B. acknowledge funding from the Fund for Scientific Research Flanders FWO project G093417N and the European Research Council under the 7th Framework Program (FP7), ERC Starting Grant 278510 VORTEX. The Qu-Ant-EM microscope used in this work was partly funded by the Hercules fund from the Flemish Government. MdH acknowledges financial support from the ANR-COSMOS (ANR-12-JS10-0002). MdH and ML acknowledge funding from the Laboratoire d'excellence LANEF in Grenoble (ANR-10-LABX-51-01).

\bibliography{mybib}{}
\bibliographystyle{plain}


\end{document}